\documentclass[a4paper,11pt]{article}
\pdfoutput=1
\usepackage{jcappub}
\usepackage[T1]{fontenc}
\usepackage{comment}

\title{\boldmath Cosmology with a Non-minimally Coupled Dark Matter Fluid\\ I. Background Evolution}

\author[a,b,c]{Samuele Silveravalle,}
\author[a,b,c,d]{Andrea Lapi,}
\author[a,b,c]{Francesco Benetti,}
\author[a,b,c]{Stefano Liberati}

\affiliation[a]{SISSA, Via Bonomea 265, 34136 Trieste, Italy}
\affiliation[b]{Institute for Fundamental Physics of the Universe (IFPU), Via Beirut 2, 34014 Trieste, Italy}
\affiliation[c]{INFN-Sezione di Trieste, via Valerio 2, 34127 Trieste, Italy}
\affiliation[d]{IRA-INAF, Via Gobetti 101, 40129 Bologna, Italy}

\emailAdd{fbenetti@sissa.it, lapi@sissa.it, ssilvera@sissa.it, liberati@sissa.it}

\abstract{We explore a cosmological model in which dark matter is non-minimally coupled to gravity at the fluid level. While typically subdominant compared to Standard Model forces, such couplings may dominate dark matter dynamics. We show that this interaction modifies the early-time Friedmann equations, driving a phase of accelerated expansion that can resolve the horizon and flatness problems without introducing additional fields. At even earlier times, the coupling to spatial curvature may give rise to a cosmological bounce, replacing the initial singularity of standard cosmology. These results suggest that non-minimally coupled dark matter could offer a unified framework for addressing both the singularity and fine-tuning problems.}

\begin{document}
\maketitle
\flushbottom

\section{Introduction}\label{sec|intro}

Our current cosmological model is based on a simple framework: a background isotropic and homogeneous spacetime, filled with matter, radiation and vacuum energy, evolving according to the laws of General Relativity. In its standard form, most matter is assumed to be cold dark matter (CDM), and vacuum energy is modeled as a cosmological constant $\Lambda$, to originate the $\Lambda$CDM model~\cite{Bull:2015stt}. In this description the Universe underwent radiation and matter-dominated phases with decelerated expansion, followed by late-time acceleration driven by $\Lambda$. Despite its success in describing large-scale phenomena, $\Lambda$CDM faces challenges across many fronts.

Evolving the equations of General Relativity backward in time leads to a spacetime singularity, where physical quantities diverge. At such high energies, a quantum theory of gravity is expected to replace classical gravity and resolve the singularity, though no current proposal has provided experimentally falsifiable predictions~\cite{Nojiri:2017ncd}.
Observations indicate that vacuum energy (or dark energy, DE) constitutes roughly 69\% of the total energy of the Universe~\cite{Carroll:2000fy,SupernovaCosmologyProject:1998vns}, but its theoretical value differs by over 120 orders of magnitude from Quantum Field Theory predictions. About 84\% of matter is thought to be CDM, consisting of massive particles that interact gravitationally and possibly very weakly via other forces, but its properties exclude them from being part of the Standard Model~\cite{Bertone:2004pz}. While alternatives to both DE~\cite{Nojiri:2017ncd,Lapi:2023plb} and CDM~\cite{Famaey:2011kh} have been proposed, in the case of CDM the overwhelming evidence strongly supports its existence~\cite{Allen2011,Rubin1980,Planck,Paraficz2016}.
The $\Lambda$CDM model is also seriously challenged by precision cosmology. In particular, the values of the Hubble parameter $H_0$ and the matter clustering parameter $S_8$ inferred from early-Universe data are in tension with those from model-independent late-time observations~\cite{DiValentino:2021izs}, while the interpretation of DE as a cosmological constant has been questioned by recent Baryon Acoustic Oscillation analyses~\cite{DESI:2024mwx}.

Finally, the $\Lambda$CDM scenario faces two major fine-tuning issues: the flatness and horizon problems. The flatness problem refers to the near spatial flatness of the Universe today, which requires the early total energy density to be fine-tuned to within one part in $10^{60}$. The horizon problem arises from the observed isotropy of the Cosmic Microwave Background (CMB), despite the regions emitting it not having been in causal contact~\cite{Guth:1980zm}, requiring then extremely uniform initial conditions. To resolve these issues, the model is typically extended with a period of accelerated expansion before radiation domination. This is generally achieved through inflation: a phase of nearly exponential expansion driven by a new, unknown scalar field slowly rolling down a flat potential~\cite{Albrecht:1982wi,Starobinsky:1980te}. Inflation ends when the field reaches the potential minimum and decays into standard particles via reheating, a process still not fully understood~\cite{Kofman:1994rk}. Inflation also helps to dilute exotic relics from beyond the Standard Model theories (e.g., magnetic monopoles), making those theories consistent with their non-observation.

The non-minimally coupled dark matter (NMC-DM) model considered in this paper is not intended to fully resolve the known issues of $\Lambda$CDM, but rather to investigate possible effects that arise when open problems in different sectors, such as dark matter, dark energy and General Relativity, are considered together. In particular, we take advantage of the elusive nature of dark matter to probe potential deviations from the standard gravitational theory. Models of dark matter range from ultralight scalar particles, such as axions or axion-like particles (ALPs)~\cite{Peccei:1977hh,Duffy:2009ig,Chadha-Day:2021szb}, to massive compact halo objects (MACHOs), such as primordial black holes~\cite{Carr:2016drx,Green:2024bam}, with a particle mass range spanning nearly 80 orders of magnitude. Yet, experimental efforts to detect dark matter through potential interactions with other Standard Model forces, in particular via the weak force in weakly interacting particles (WIMPs) dark matter models~\cite{Jungman:1995df,Bertone:2004pz}, have only succeeded in placing an upper limit on their cross-section at $\sigma\sim 10^{-47}\mathrm{cm}^2$~\cite{XENON1T,PANDAX,LZ}. This leaves open the possibility that the first relevant interaction between dark matter particles occurs through a modified gravitational coupling.

Non-minimal couplings between matter fields and curvature are commonly introduced when considering quantum fields in curved spacetimes, as they emerge from the renormalization group flow~\cite{Birrell:1982ix}, or in inflationary models as they naturally give rise to effective potentials for the scalar field~\cite{Hertzberg:2010dc}. In other words, non-minimal couplings are the expected deviations from the classical gravitational interaction of matter at high energies. Nonetheless, in this paper we consider a model in which the non-minimal coupling acts on dark matter at the fluid level, rather than at the field level, in order to avoid assuming a specific dark matter model and, at the same time, to account for the possibility of the NMC being an emergent phenomenon~\cite{Bettoni14}. This idea has been developed in recent years, and the general properties of a NMC-DM fluid~\cite{Bettoni11,Bettoni12,Bettoni15}, along with the consequences of its Newtonian limit on cosmological and galactic scales~\cite{Gandolfi21, Gandolfi22,Gandolfi23}, have been investigated.

In a recent paper, we presented a fully relativistic realization of this model in the context of static and spherically symmetric configurations~\cite{Benetti_2025}. The non-minimal interaction can be interpreted as an effective stress-energy tensor for dark matter, allowing a fluid modeled as collisionless dust to develop an effective pressure capable of supporting self-gravitating dark matter configurations. These configurations exhibit very high mass-to-radius ratios and develop light ring structures; hence, they have been named non-minimally coupled ultra-compact objects (NMC UCOs). The scales of these solutions are fully determined by a single parameter, which has been fixed by assuming a cross section smaller than any found in the Standard Model~\cite{CMS:2018ccg,CMS:2020fqz}. This choice leaves open the possibility that the non-minimal coupling is not exclusive to dark matter, but may be a property shared by all known particles. To link the cross section of the interaction with the macroscopic properties of NMC UCOs, we considered a threshold density at which the mean free path of particles $\lambda_{\text{mfp}}=m_\chi/\rho_\chi\sigma_{\chi\chi}$ coincides with the curvature radius of the solution $R_c=c(8\pi G \rho_\chi)^{-1/2}$:
\begin{equation}\label{eq|thresdens}
    \rho_{\text{th}}=\frac{8\pi G}{c^2}\left(\frac{\sigma_{\chi\chi}}{m_{\chi}}\right)^{-2};
\end{equation}
with a particle mass $m_\chi\sim 100$ GeV and a cross section $\sigma_{\chi\chi}\sim 10^{-37}\mathrm{cm}^2$, this gives a threshold density of $\rho_{\text{th}}\sim 10^4\mathrm{g}/\mathrm{cm}^3$ and UCO masses of the order of $M\sim 10^5 M_\odot$, compatible with the ones expected for a supermassive black hole seed~\cite{Johnson:2012cw}. In general, deviations from GR are expected to scale with the ratio $\rho_{\rm DM}/\rho_{\rm th}$. This corresponds to deviations of order $10^{-28}$ in the Solar System, and $10^{-20}$ at the Galactic center, assuming that DM follows a NFW profile. The NMC under consideration is therefore not expected to have significant consequences at these scales and should satisfy all current constraints.

While the required densities are much smaller than those of other compact and ultra-compact objects, they may still be difficult to achieve through the collapse of collisionless dark matter in the late universe. For this reason, it is sensible to investigate the role of NMC-DM in a cosmological context, as a possible environment for the formation of UCOs. In addition, it is necessary to examine this cosmological model to determine whether the non-minimal coupling contradicts observational data, which would render the model unphysical. As we will show in the following sections, cosmology with an NMC-DM fluid actually moves in the direction of addressing some of the problems of the $\Lambda$CDM model, naturally providing an inflationary phase and the possibility of avoiding the Big Bang singularity through a cosmological bounce.

\section{Non-minimally coupled dark matter}\label{sec|NMCDM}

As stated in the introduction, non-minimal couplings are usually introduced either as high-energy corrections, motivated by quantum field theory arguments~\cite{Birrell:1982ix,Hertzberg:2010dc}, or as emergent phenomena related to the role of spacetime curvature at scales where DM can be described as a fluid~\cite{Bettoni11,Bettoni15}, for example when DM is modeled as a Bose–Einstein condensate with a sufficiently large healing length~\cite{Bettoni14}. In this paper we do not assume a specific mechanism generating the non-minimal coupling, but simply that its effects can extend up to scales where DM behaves as pressureless dust. Such a phenomenological approach reduces, at the practical level, to considering a non-minimally coupled DM fluid~\cite{Bettoni15}. 
Including the Lagrangian of other minimally coupled forms of matter, the total action of the theory reads
\begin{equation}\label{eq|action}
    S_{\rm NMC} = S_{\rm MC} + S_{\rm int} = \int_{\mathcal{M}}{\rm d}^4x\; \sqrt{-g}\,\left[\frac{c^4}{16\pi\,G}R + \mathcal{L}_{\rm mat} + \mathcal{L}_{\rm DM}\right]+\int_{\mathcal{M}}{\rm d}^4x\; \sqrt{-g}\,\epsilon L^2 G_{\mu\nu}\,T^{\mu\nu}_{\rm DM},
\end{equation}
where $\mathcal{L}_{\rm mat}$ denotes the Lagrangian for all non-DM components, and in the interaction term $T^{\mu\nu}_{\rm DM}$ is taken to be $T^{\mu\nu}_{\rm DM} = \rho \,c^2\,u^{\mu}\,u^{\nu}$, with $\rho$ the DM mass-energy density and $u^{\mu}$ its four-velocity. We stress that the choice of our NMC term is not guided by a standard EFT argument, but rather by the (weak) equivalence principle. A detailed reasoning can be found in section 2 of~\cite{Bettoni11} and is based on the findings of Bekenstein in~\cite{Bekenstein93}. In particular, the specific interaction we adopt is motivated by previous work on BEC dark matter, where the condensation mechanism provides a long range coherence length which allows to construct the Horndeski type NMC term we need~\cite{Bettoni:2013diz}. With such a coupling, the equations of motion in the spherically symmetric case can still be recast as two first-order equations~\cite{Benetti_2025}, and also in the cosmological case the differential order of the equations is not modified, as we will show in the next section.

The interaction Lagrangian features a characteristic length scale $L$, required for the correct dimensionality, and a dimensionless `polarity' parameter $\epsilon = \pm\,1$. These parameters can be seen as the macroscopic manifestation of the microphysics of the coupling between dark matter and curvature, of whose origin we remain agnostic in this work. In~\cite{Benetti_2025} we have found that, in the context of NMC UCOs, only the choice $\epsilon=-1$ leads to regular configurations, while the opposite choice $\epsilon=+1$ always results in singular or non-asymptotically flat solutions. 
We also found that the mass of the solutions is proportional to $L$, while the central density scales as $L^{-2}$; moreover, there exists a minimum density below which no non-trivial solution can be found.
By equating this minimum density with the threshold density in~\eqref{eq|thresdens}, we determined a value of $L \simeq 3.7\times10^{11}\,\rm cm$, corresponding to masses of the order of $10^{5}\,M_{\odot}$ for NMC-UCOs. To maintain consistency with our previous work, we adopt $L \simeq 3.7\times10^{11}\,\rm cm$ as the fiducial value, set $\epsilon=-1$ and $c=1$.

The gravitational field equations of the model are derived by varying the action~\eqref{eq|action} with respect to the metric. In doing so one has to recall the variation of the Einstein tensor and of the stress-energy tensor of a perfect fluid as reported in~\cite{haghani2023variation}, 
\begin{align}
 \delta G_{\mu\nu} = &\  \nabla^{\alpha}\nabla_{(\mu}\,\delta g_{\nu)\alpha}-\frac{1}{2}\Box\,\delta g_{\mu\nu} -\frac{1}{2}g^{\alpha\beta}\nabla_{(\mu}\nabla_{\nu)}\,\delta g_{\alpha\beta}  - \frac{1}{2}R\,\delta g_{\mu\nu}+ \nonumber \\
 & - \frac{1}{2}\,g_{\mu\nu}\,(R^{\alpha\beta} + g^{\alpha\beta}\,\Box - \nabla^{\alpha}\,\nabla^{\beta})\,\delta g_{\alpha\beta}\,,\\
\delta T^{\mu\nu}_{\rm DM} = &\ \frac{1}{2}T^{\mu\nu}_{\rm DM}\,(u^{\alpha}u^{\beta} - g^{\alpha\beta})\,{\delta g_{\alpha\beta}}\,.
\end{align} 
Then the field equations for the gravitational field sourced
by the non-minimally coupled DM fluid read
\begin{align}\label{eq|fieldeq}
    \frac{1}{8\pi\,G}\,G^{\mu\nu}& = T_{\rm mat}^{\mu\nu} + T_{\rm DM}^{\mu\nu} - L^2\,\left[(G^{\mu\nu} + g^{\mu\nu}\,\Box - \nabla^{(\mu}\nabla^{\nu)})\,T_{\rm DM} - \Box\,T_{\rm DM}^{\mu\nu} + 2\,\nabla^{\alpha}\,\nabla^{(\mu}\,T^{\nu)}_{\alpha} \right. \nonumber \\
    & \left.- g^{\mu\nu}\nabla_{\alpha}\,\nabla_{\beta}\,T^{\alpha\beta}_{\rm DM} - \frac{R}{2}\,(\,T_{\rm DM}^{\mu\nu}-g^{\mu\nu}\,T_{\rm DM}) - \frac{T_{\rm DM}^{\mu\nu}}{T_{\rm DM}}\,R_{\alpha\beta}\,T^{\alpha\beta}_{\rm DM}\right]\, .
\end{align}
Although this procedure is relatively simple, it nonetheless ensures that the constraints on the fluid dynamics required for the perfect fluid approximation, namely particle number conservation and absence of entropy exchanges between flow lines, are satisfied~\cite{Bettoni15}.

The presence of the non-minimal coupling indicates a deviation from the pressureless fluid approximation within a perturbative framework, where the fluid can in principle develop pressure, stresses, and energy fluxes whenever its density or the spacetime curvature become sufficiently large. Similar effects appear in effective theories of fluids~\cite{Dubovsky:2011sj,Ballesteros:2012kv,Ballesteros:2014sxa}, where higher derivatives of the fluid degrees of freedom are included as high-energy corrections to the perfect fluid approximation. These corrections can mimic modified-gravity behaviors~\cite{Ballesteros:2016gwc,Berezhiani:2015bqa}, thus bridging DM and modified gravity theories, as we do in our model. While exploring possible links with generalized fluid dynamics would be of great interest in the future, our framework already captures a wide range of processes, from fundamental modifications of gravitational interactions to alternative fluid descriptions in curved spacetimes, while also taking into account possible self-interactions through the specific coupling with the Einstein tensor.

\subsection{Effective stress-energy tensor and modified gravity}

By looking at Equation~\eqref{eq|fieldeq} one can see that in the context of the present model, gravity is generated not only by the DM energy density, but also by its time and space variations, implying that the minimally-coupled DM stress-energy tensor $T^{\mu\nu}_{\rm DM}$ is no longer conserved. This can be seen explicitly by taking the covariant divergence of equation~\eqref{eq|fieldeq} and using the Bianchi identities
\begin{align}\label{eq|Bianchi}  
\nabla_{\mu}T^{\mu\nu}_{\rm mat} + \nabla_{\mu}T^{\mu\nu}_{\rm DM} &+ L^2\,\left\{ R^{\mu\nu}\nabla_{\mu}\,T_{\rm DM} + \nabla^{\nu}R_{\mu\lambda}\,T^{\mu\lambda}_{\rm DM} -2\nabla_{\mu}(R^{\nu}_{\lambda}\,T^{\lambda\mu}_{\rm DM}) \nonumber \right.\\
& \left.+\nabla_{\mu}\left[\,\frac{R}{2}\,(T^{\mu\nu}_{\rm DM}-g^{\mu\nu}\,T_{\rm DM}) + \frac{T^{\mu\nu}_{\rm DM}}{T_{\rm DM}}\,R_{\alpha\beta}\,T^{\alpha\beta}_{\rm DM}\,\right]\right\} = 0\,.
\end{align}
While the stress-energy tensor of ordinary matter can be separately conserved $\nabla_\mu T^{\mu\nu}_{\rm mat}=0$, if we assume that there are no direct interactions with DM, one can see that the DM stress-energy tensor satisfies a more involved expression.

However, we can interpret Equation~\eqref{eq|fieldeq} as the Einstein field equations of General Relativity sourced by an effective, or non-minimally coupled, DM stress-energy tensor $G^{\mu\nu} = 8\pi G(T_{\rm mat}^{\mu\nu} + T_{\rm DM, eff}^{\mu\nu})$~\cite{Bettoni12,Bettoni15}. Then Equation~\eqref{eq|Bianchi} expresses the conservation of the effective stress-energy tensor, while the NMC with gravity prevents the minimally-coupled $T_{\rm DM}^{\mu\nu}$ to be separately conserved. The effective energy density and pressures associated to $T_{\rm DM, eff}^{\mu\nu}$ have a dynamical nature, meaning that they emerge as the result of the non-minimal coupling of the original DM fluid with the gravitational field; as such, they depend on the particular form of the metric solution of the field equations. Once a metric has been specified, the values of the energy density and pressures can be found by comparison of~\eqref{eq|fieldeq} with Einstein equations. In particular, in~\cite{Benetti_2025} we have shown that the emergence of a radial effective pressure allows the collisionless DM fluid to settle in spherically symmetric, ultra-compact hydrostatic equilibrium configurations. 

However, while in the context of NMC UCOs, where only dark matter is present, the interpretation of Equation~\eqref{eq|fieldeq} in terms of an effective stress-energy tensor is straightforward, this interpretation becomes less trivial in a cosmological context, where baryons, radiation, and dark energy are also present. Since the effective stress-energy tensor is metric-dependent, and the evolution of the metric depends on the total matter content of the universe, Equation~\eqref{eq|Bianchi} implies an indirect interaction between dark matter and the other energy components through gravity. It is therefore possible to move the NMC term in Equation~\eqref{eq|fieldeq} to the left-hand side and interpret it as a modification to gravity. Our NMC-DM model is designed precisely to bridge the gap between dark matter models and modified gravity theories, and it is thus natural that both interpretations are applicable. In the next section, we will show that this interpretation is particularly well suited in a cosmological context, as the conservation equation for dark matter remains unchanged, while the Friedmann equations are modified.

\section{Non-minimally coupled cosmology}

To look for implications of the non-minimal coupling in a cosmological context, we model the spacetime with a Friedmann-Lemaître-Robertson-Walker (FRLW) metric
\begin{equation}\label{eq|friedmannmetric}
ds^2 = -dt^2+a^2(t)\left(\frac{dr^2}{1-k\,r^2}+r^2d\Omega^2\right);
\end{equation}
here $t$ is the cosmic time measured by comoving observers, $a(t)$ is the scale factor, and $k=0,\pm R_0^{-2}$ is the Gaussian curvature of space at present time. The matter content of the model comprises four comoving and homogeneous perfect fluids: baryons, cold dark matter, radiation (photons and neutrinos), and dark energy (in the form of a cosmological constant $\Lambda$). The equations of state for these cosmic fluids have the form $p_i(t) = w_i\rho_i(t)$, with $w_b = w_{\rm DM} = 0, \, w_{r} = 1/3, \, w_{\Lambda} = -1$.

The only two non-vanishing components of the field equations~\eqref{eq|fieldeq} with the ansatz~\eqref{eq|friedmannmetric} are the time-time component and a combination of the latter with the trace
\begin{subequations}\label{eq|friedmann}
\begin{equation}
    H^2 = \frac{8\pi G}{3}\frac{\rho_{\rm DM} + \rho_{b} + \rho_{r} + \rho_{\Lambda}}{1+8\pi G L^2\rho_{\rm DM}} - \frac{k}{a^2}\frac{1-8\pi G L^2\rho_{\rm DM}}{1+8\pi G L^2\rho_{\rm DM}}, \label{eq|friedmann00}
\end{equation}
\begin{equation}
    \dot H+H^2= -\frac{4\pi G}{3} \frac{\left(1-9 L^2 \left(H^2-\frac{k}{a^2}\right)\right)\rho _{\rm DM}+\rho_b + 2 \rho_r-2\rho _{\Lambda}}{1+8 \pi  G L^2\rho_{\rm DM}}.\label{eq|Raychaudhuri}
\end{equation}
\end{subequations}
Equation~\eqref{eq|friedmann00} is called the Friedmann equation, while~\eqref{eq|Raychaudhuri} is the Raychaudhuri equation.
The time component of~\eqref{eq|Bianchi} yields instead the continuity equation
\begin{equation}\label{eq|conservation}
\left[1+3 L^2\left(H^2+\frac{k}{a^2}\right)\right]\left(\dot\rho_{\rm DM}+3H\rho_{\rm DM}\right)+\dot\rho _b+3H\rho_b + \dot\rho _r+4H\rho_r + \dot\rho _{\Lambda }=0.
\end{equation}
If there are no energy transfers between different components of the cosmic fluid, then Equation~\eqref{eq|conservation} splits into the four conditions $\dot\rho_i + 3H(1+w_i)\rho_i=0$, one for each component, with solution $\rho_i(t) = \rho_{i,0}\,a(t)^{-3(1+w_i)}$. Here the subscript ``${}_0$'' refers to a quantity evaluated at the present time, and $a_0$ has been set equal to one. Note that with the FLRW ansatz for the metric the additional terms due to the non-minimal coupling with gravity factorize, so that DM behaves like dust.

In the context of cosmology it is customary to express the evolution of quantities in term of redshift $1+z = a^{-1}$, and to define the dimensionless density parameters 
\begin{equation}\label{def|Omegai}
\begin{split}
    & \Omega_i \equiv \frac{8\pi G}{3H^2}\rho_i, \\
    & \Omega_k \equiv -\frac{k}{a^2H^2},
\end{split}
\end{equation}
related to their present values by 
\begin{equation}\label{Omegaevolution}
\begin{split}
    & \Omega_i = (H_0/H)^2\,\Omega_{i,0}\,(1+z)^{3(1+w_i)}, \\
    & \Omega_k = (H_0/H)^2\,\Omega_{k,0}\,(1+z)^{2}. 
\end{split}
\end{equation}
Note that the evolution of $\Omega_k$ suggests the analogy with a fluid with an equation of state parameter $w_k = -1/3$. To keep track of the relevance of the non-minimal coupling during the evolution of the Universe, we also define the quantity
\begin{equation}
    \chi \equiv 8\pi G L^2\rho_{\rm DM} = 3 L^2H^2\Omega_{\rm DM},
\end{equation}
evolving as $\chi = \chi_0\,(1+z)^{3}$. We would like to underline that $\rho_{\rm DM} = 1/8\pi G L^2$, i.e. $\chi = 1$, corresponds precisely to the critical density at which NMC UCOs develop a central singularity. However, in the cosmological context there is no divergence in the curvature invariants at this value, and the evolution is regular at any finite redshift. We also stress that a singularity would indeed appear at $\chi = 1$ if we had chosen the polarity parameter $\epsilon = +1$, which would reverse the sign of all terms multiplied by $L^2$. This supports our decision to avoid $\epsilon = +1$, as it leads to singular configurations in both cosmological and astrophysical contexts.

With the above notation, Equation~\eqref{eq|friedmann00} can be written as a constraint involving the sum of the density parameters
\begin{equation}\label{constraint}
    \Omega_m + \Omega_r + \Omega_\Lambda + \Omega_k\,(1-\chi) = 1 + \chi,
\end{equation}
where $\Omega_{m} \equiv \Omega_{\rm DM} + \Omega_{b}$ denotes the total matter density parameter. Taking into account~\eqref{Omegaevolution}, we can further write Equation~\eqref{eq|friedmann00} explicitly as a function of redshift
\begin{equation}\label{eq|friedmannchi}
 \left(\frac{H}{H_0}\right)^2 = \frac{\Omega_{m,0}(1+z)^3 + \Omega_{r,0}(1+z)^4 + \Omega_{\Lambda,0} + \Omega_{k,0}(1+z)^2\,(1-\chi_0(1+z)^3)}{1+\chi_0(1+z)^3}.
\end{equation}
In the calculations presented in the next subsections we use values of the density parameters and of the Hubble constant today from the Planck collaboration~\cite{Planck}, and $\chi_0 = 5.85\times 10^{-34}$ given by our fiducial value of $L$.

\subsection{Cosmic evolution}\label{sec|cosmevo}

A first grasp of the evolution of the Universe within this model can be attained by looking at equation~\eqref{eq|friedmannchi} at different epochs, starting from the present time and going towards higher redshifts. First of all, one notes that the non-minimal coupling of DM with gravity affects the evolution of the Universe only at early times when $\chi\geq 1$, while it can be completely neglected at later times (at least at the background level) when the expansion of the Universe causes $\chi$ to drop significantly below one. Thus, the late time evolution will be indistinguishable from the one in the standard $\Lambda$CDM paradigm
\begin{equation}\label{eq|LCDM}
     \left(\frac{H}{H_0}\right)^2 = \Omega_{m,0}(1+z)^3 + \Omega_{r,0}(1+z)^4 + \Omega_{\Lambda,0} + \Omega_{k,0}(1+z)^2,
\end{equation}
with a DE dominated phase extending to redshift $\log(1+z) \simeq 0.11$, a matter dominated epoch in the range $0.11 \leq \log(1+z) \leq 3.54$, and radiation domination at higher redshifts. As the DM density increases with redshift, a point will be reached when $\chi\simeq 1$ and at earlier times, even though radiation is still the dominant component, the NMC of DM with gravity causes the evolution to depart from the $\Lambda$CDM scenario. The relevant redshift is given by
\begin{equation}
    \log\left(1+z_{\rm NMC}\right) \equiv -\frac{1}{3}\log\left(3\,\Omega_{\rm DM,0}\right) -\frac{2}{3}\log\left(H_0\,L\right) \simeq 11.08.
\end{equation}
We note that the chosen value of $L$ ensures that the NMC becomes irrelevant before neutrino decoupling, and thus earlier than any observable phenomenon.

We can get an intuition about the evolution in the NMC-dominated era by looking at Equation~\eqref{eq|friedmannchi} in the deep limit $\chi\gg 1$, and neglecting matter and DE, since they are subdominant at such high redshifts
\begin{equation}\label{eq|highz}
    \left(\frac{H}{H_0}\right)^2 =\frac{ \Omega_{r,0}}{\chi_0}(1+z)- \Omega_{k,0}(1+z)^2.
\end{equation}
For a spatial curvature parameter close to its limiting value $|\Omega_{k,0}|\sim10^{-3}$, the first term on the right hand side of~\eqref{eq|highz} dominates until a redshift
\begin{equation}
    \log\left(1+z_{k}\right) \equiv \log\left(\frac{\Omega_{r,0}}{3\,|\Omega_{k,0}|\,\Omega_{\rm DM,0}}\right)-2\log\left(H_0\,L\right) \simeq 32.20,
\end{equation} 
leading to an almost quadratic evolution of the scale factor during this period
\begin{equation}\label{eq|infevo}
    a(t) = a(t_k)\left(1+\frac{H_0}{2}\sqrt{\frac{\Omega_{r,0}}{a(t_k)\chi_0}}(t-t_k)\right)^2.
\end{equation}

At even higher redshifts $z \gg z_k$, the evolution of the Universe strongly depends on the sign of the spatial curvature. In the case of a vanishing curvature $\Omega_k=0$, the Universe is dominated by radiation and follows the evolution in equation~\eqref{eq|infevo} at all times.
On the other hand, even if a small positive curvature (corresponding to $\Omega_k<0$) is present, at very early time it would be the dominant component, yelding a linear expansion of the Universe ($a(t) \propto \sqrt{k}t)$. One can thus conclude that both a positive or vanishing curvature imply the existence of a singularity in the remote past, although attained at different rates, quadratic in the first case, linear in the second. The most interesting case however, is that of a negative spatial curvature ($\Omega_k>0$). Looking at Equation~\eqref{eq|highz}, we see that in this scenario the expansion rate vanishes at $z_k$ signaling the presence of a cosmological bounce. It is interesting to note that, for the chosen value of $L$ and the limiting value of spatial curvature $|\Omega_{k,0}|\sim10^{-3}$, the bounce would have occurred when the radiation temperature was extremely close to the Planck temperature $T_P\sim10^{32}K$. 

\begin{figure}[t!]
    \centering
    \includegraphics[scale=0.7]{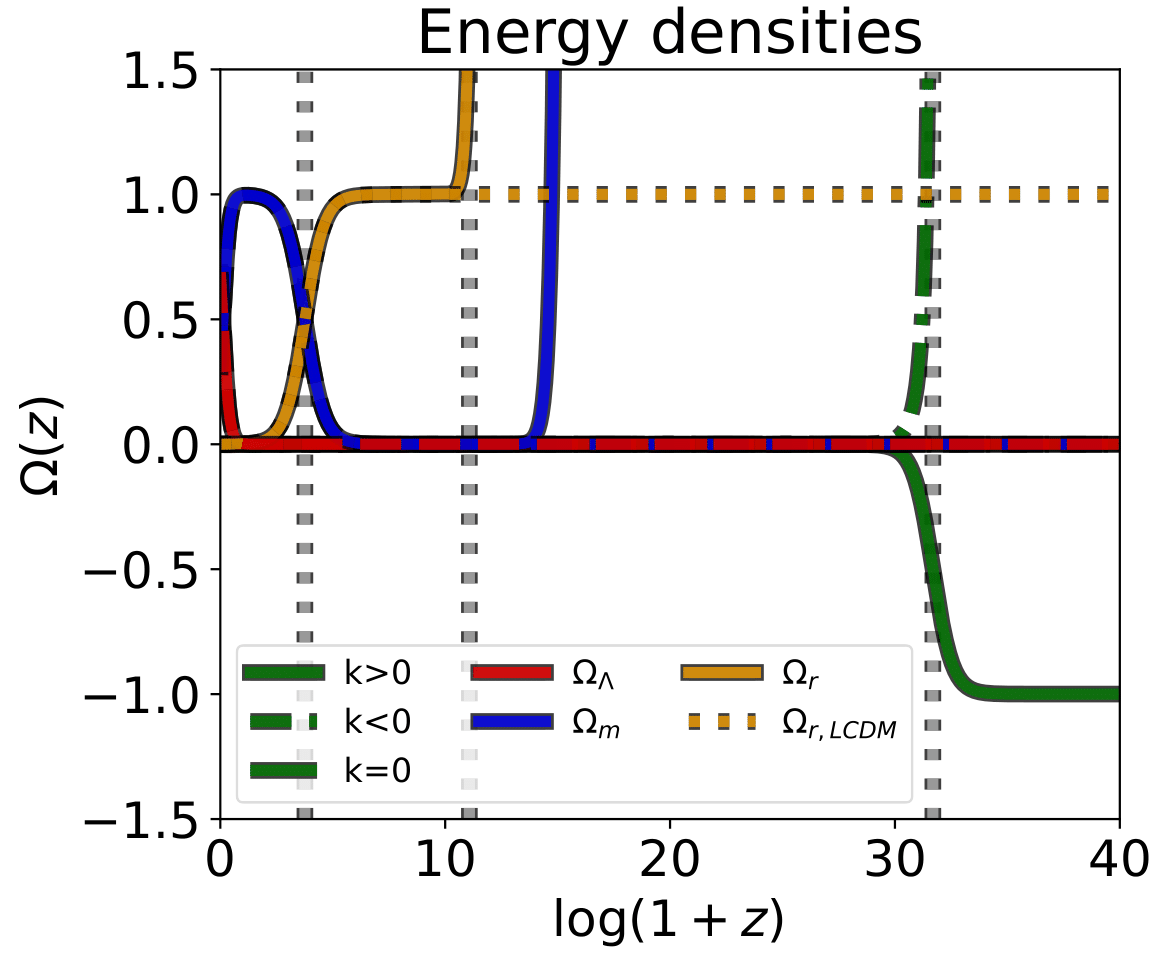}
    \caption{Evolution of the density parameters as a function of redshift $1+z \equiv a^{-1}$. Solid lines represent the solution in the present non-minimal coupling model, while the dotted lines show the corresponding ones in the $\Lambda \rm CDM$ model. The vertical dotted grey lines represent matter-radiation equality at $\log\left(1+z_{\rm eq}\right) \simeq 3.54$, the onset of the NMC at redshift $\log\left(1+z_{\rm NMC}\right)\simeq 11.08$, and the beginning of the curvature era (or the redshift of the bounce), at redshift $\log\left(1+z_{k}\right) \simeq 32.20$.}  
    \label{fig|cosmo}
\end{figure}

Figure~\ref{fig|cosmo} shows the evolution of the density parameters~\eqref{def|Omegai} as a function of redshift. Solid lines represent the NMC solution, while the dotted lines show the corresponding solution in the $\Lambda \rm CDM$ model.  The vertical dotted grey lines represent matter-radiation equality at $\log\left(1+z_{\rm eq}\right) \simeq 3.54$, the onset of the NMC at redshift $\log\left(1+z_{\rm NMC}\right)\simeq 11.08$, and the beginning of the curvature era (or the redshift of the bounce), at redshift $\log\left(1+z_{k}\right) \simeq 32.20$. The green lines indicate the quantity $\Omega_k\frac{1-\chi}{1+\chi}$ with $k\lesseqgtr0$. From this plot it is clear that radiation remains the dominant component of the cosmic fluid even during the NMC-dominated phase, while at the same time the dynamics doe not follow any longer the standard Friedmann equation, as the constraint equation~\eqref{constraint} does not longer enforces the density parameters to sum to one.

\begin{figure}[t!]
    \centering
    \includegraphics[scale=0.7]{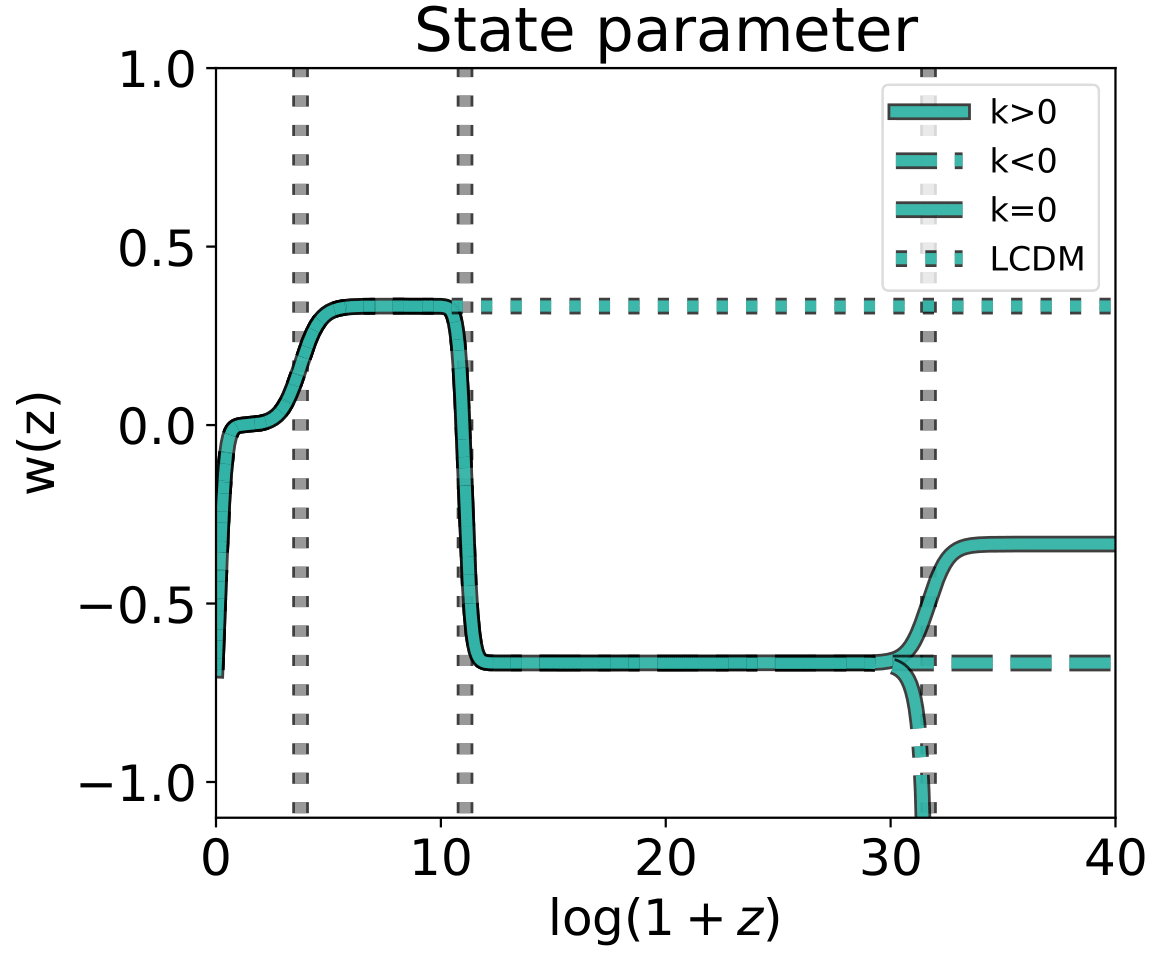}
    \caption{The dynamical state parameter $w$ as a function of redshift. The solid, dashed and dotted-dashed lines are the behavior in the presence of NMC with negative, null, or positive curvature respectively, while the dotted line is the one in $\Lambda$CDM. The vertical dotted grey lines represent matter-radiation equality at $\log\left(1+z_{\rm eq}\right) \simeq 3.54$, the onset of the NMC at redshift $\log\left(1+z_{\rm NMC}\right)\simeq 11.08$, and the beginning of the curvature era (or the redshift of the bounce), at redshift $\log\left(1+z_{k}\right) \simeq 32.20$.}
    \label{fig|w}
\end{figure}

To keep track of the roles played by the different fluid components during the evolution of the Universe, it is useful to define a dynamical state parameter
\begin{equation}\label{effw}
    w \equiv -\left(1+\frac{2}{3}\frac{\dot H}{H^2}\right),
\end{equation}
so that at any time in which $w$ is almost constant, the scale factor evolves as
\begin{equation}
    a(t) \simeq \left(\frac{t}{t_0}\right)^{\frac{2}{3(1+w)}}.
\end{equation}
In the absence of NMC, when a fluid component dominates the energy content of the Universe,~\eqref{effw} coincides with the state parameter of that component. In the presence of the NMC between DM and gravity this correspondence is more involved. Substituting Equation~\eqref{eq|friedmann} in~\eqref{effw}, we find 
\begin{equation}
    w= \frac{\Omega_{m} + 2\,\Omega_r -2\,\Omega_{\Lambda} -3\,\chi\left(1+\Omega_k\right)}{3\,\left(1+\chi\right)} -\frac{1}{3}.
\end{equation}
When $\chi \ll 1$ one has the familiar values $w=-1$, $w=0$, $w=1/3$ during DE, matter and radiation domination respectively. In the deep NMC regime $\chi \gg 1$, even if radiation is still the dominant component, the constraint~\eqref{constraint} requires $\Omega_r \simeq \chi$, leading to $w = -2/3$, and to an accelerated quadratic expansion. At very high redshift, in the case of a positive curvature $\Omega_{k,0}<0$, the curvature dominates the energy density and the effective parameter increases to $w=-1/3$. The evolution of the dynamical state parameter as a function of redshift is shown in Figure~\ref{fig|w}. The solid, dashed and dot-dashed lines are the behavior in the presence of NMC with negative, null, or positive curvature respectively, while the dotted line is the one in $\Lambda$CDM. As mentioned, this plot has two peculiar features. First, one sees a steep decrease of the state parameter at the onset of NMC, when $\chi\simeq 1$, leading to an accelerated phase of expansion at higher redshifts. We want to stress that during this epoch radiation is still the dominant energy component, meaning that $\rho_r \gg \rho_m \gg \rho_{\rm DE}$, but the dynamical state parameter departures from the value of $w_r=1/3$, characteristic of radiation, because of the non-trivial coupling between DM and gravity. Furthermore, if a curvature is present, at redshift $z_k$, it becomes the dominant component, and either the state parameter increases to $w_k=-1/3$, characterizing curvature domination, or the Universe has a bounce and the state parameter is no longer well defined.

\subsection{NMC-driven inflation and bounce}

\begin{figure}[t!]
    \centering
    \includegraphics[scale=0.7]{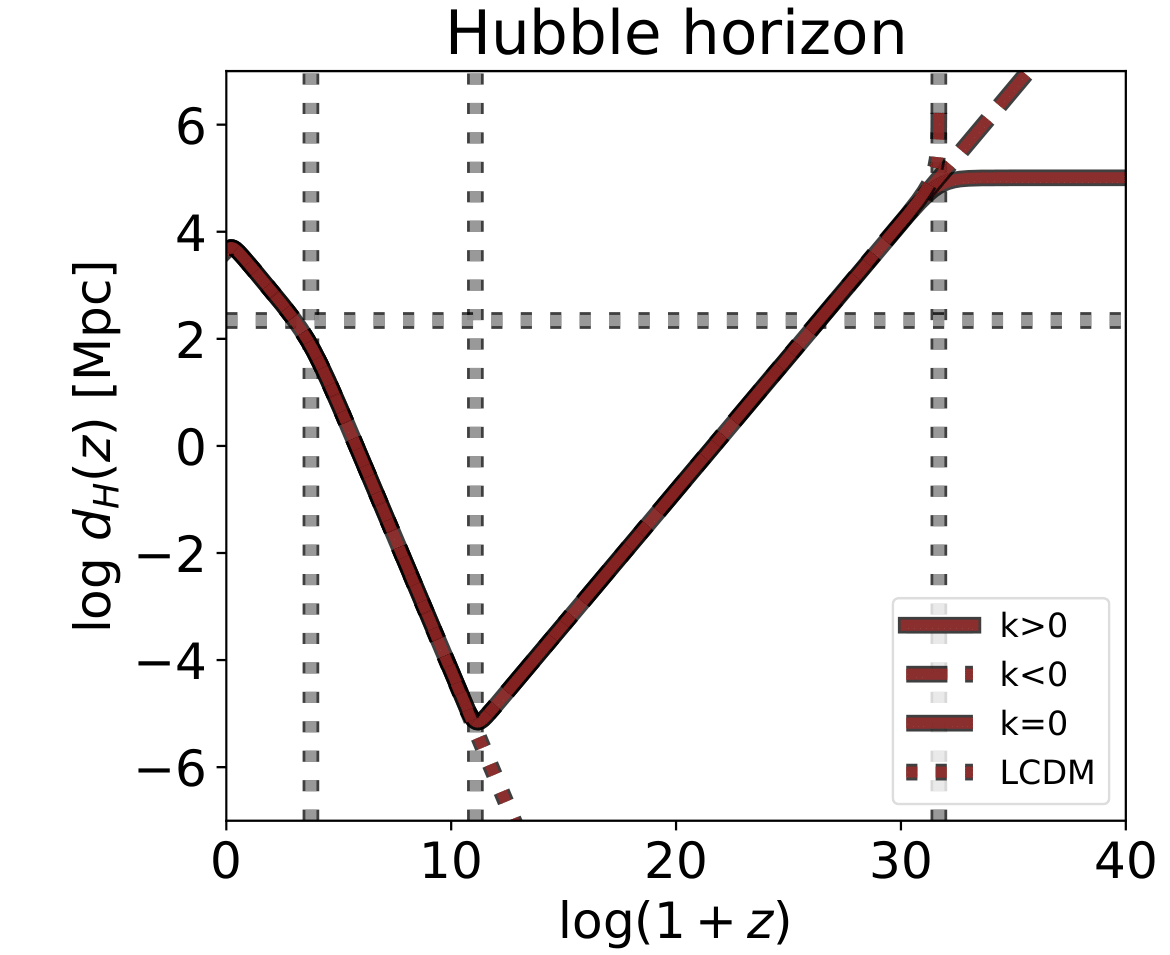}
    \caption{The comoving Hubble horizon as a function of redshift. The solid, dashed and dotted-dashed lines are the behavior in the presence of NMC with negative, null, or positive curvature respectively, while the dotted line is the one in $\Lambda$CDM. The vertical dotted grey lines represent matter-radiation equality at $\log\left(1+z_{\rm eq}\right) \simeq 3.54$, the onset of the NMC at redshift $\log\left(1+z_{\rm NMC}\right)\simeq 11.08$, and the beginning of the curvature era (or the redshift of the bounce), at redshift $\log\left(1+z_{k}\right) \simeq 32.20$. The horizontal dashed line shows the value of the horizon at the epoch of recombination at $\log\left(1+z_{\rm CMB}\right) \simeq 3.04$. The line further intersects the horizon at $\log\left(1+z_{\rm end}\right) \simeq 26.36$.}
    \label{fig|horizon}
\end{figure}

As opposed to standard inflationary scenarios, which require the introduction of an unknown scalar field, the NMC model features an accelerated epoch that emerges naturally from the non-minimal coupling between dark matter and gravity, once $\chi\geq 1$. This accelerated phase, during which the comoving Hubble horizon decreases with time, offers the possibility of solving the horizon and flatness problems without introducing new components to the energy content of the Universe.
Figure~\ref{fig|horizon} shows the evolution of the comoving Hubble horizon, defined as
\begin{equation}
    d_H(z) = \frac{1}{a(z)H(z)},
\end{equation}
as a function of redshift. 
In order to solve the horizon problem, this phase of accelerated expansion must last long enough for the comoving Hubble horizon to grow from its minimum value at the onset of radiation domination to the value it has at recombination, around redshift $z_{\rm CMB}\simeq 1100$. The flatness problem is commonly described in terms of a critical total energy density that would make the Universe spatially flat. In General Relativity, one has that
\begin{equation}\label{eq|flatness}
    \rho_c=\frac{3H^2}{8\pi G},\qquad\implies\qquad\frac{\rho_m+\rho_r+\rho_\Lambda}{\rho_c}-1=\frac{k}{a^2 H^2}=-\Omega_k=d_H^2k,
\end{equation}
implying that to ensure $|\Omega_k|<10^{-3}$ today, the total energy density in the early Universe must have satisfied $\rho\sim\rho_c(1\pm10^{-60})$, which is extremely fine-tuned. In the NMC model, however, equation~\eqref{eq|flatness} is modified as follows:
\begin{equation}\label{eq|flatnessnmc}
    \rho_c=\frac{3H^2(1+\chi)}{8\pi G},\qquad\implies\qquad\frac{\rho_m+\rho_r+\rho_\Lambda}{\rho_c}-1=\frac{k}{a^2 H^2}\frac{1-\chi}{1+\chi}=d_H^2k\frac{1-\chi}{1+\chi};
\end{equation}
that is, the deviation from flatness is now proportional to $d_H^2k\frac{1-\chi}{1+\chi}$ instead of $d_H^2k$. Nonetheless, it is evident that solving the horizon problem also alleviates the flatness problem, since the deviation from the critical density remains proportional to $d_H^2$, except during the short period where $\chi\sim 1$, which, however, does not significantly affect the overall evolution.
In Figure~\ref{fig|horizon} the horizontal dotted line shows the value of the comoving Hubble horizon at recombination. The line further intersects the horizon at $\log\left(1+z_{\rm end}\right) \simeq 26.54$, and therefore to solve the horizon problem the accelerated expansion had to start before $z_{\rm end}$, and then to last a minimum of $\log\left(1+z_{\rm end}\right)-\log\left(1+z_{\rm NMC}\right) \simeq 15.46$. 
Considering $z_{k}$ as the beginning of the accelerated phase, we thus find
\begin{equation}
    \log\left(1+z_{k}\right)-\log\left(1+z_{\rm NMC}\right) = \log\left(\frac{\Omega_{r,0}}{|\Omega_{k,0}|\,\left(3\,\Omega_{\rm DM,0}\right)^{2/3}}\right) - \frac{4}{3}\log(H_0\,L) \gtrsim 15.46,
\end{equation}
which implies
\begin{equation}
    |\Omega_{k,0}|\lesssim \frac{\Omega_{r,0}}{(3 \Omega_{\rm DM,0} H_0^2L^2)^{2/3}}10^{-15.46}=\Omega_{r,0}\chi_0^{-2/3}10^{-15.46}\simeq 4.56\times 10^2,
\end{equation}
that is clearly well guaranteed by the Planck collaboration constraint $|\Omega_{k,0}|< 1.2\times10^{-3}$. 
Thus, as it is evident from Figure~\ref{fig|horizon}, the epoch of accelerated expansion provided by the NMC between DM and gravity is sufficient to solve the horizon problem. In particular, with the parameters considered we find $\log\left(1+z_{k}\right)-\log\left(1+z_{\rm NMC}\right) \simeq 21.12$, which would correspond to approximately $N=48.62$ $e$-folding. 

As for the beginning of the Universe, we would like to review the possibility of a cosmological bounce at redshift $z_k$ with negative spatial curvature. It is clear from the structure of Equation~\eqref{eq|highz} that, for positive $\Omega_{k,0}$, higher redshift values are not permitted. However, the absence of divergences in the curvature invariants indicates that the spacetime can be extended, implying that the scale factor must begin to increase again. Equation~\eqref{eq|highz} can indeed be integrated for positive $\Omega_{k,0}$, and is solved by
\begin{equation}\label{eq|bounce}
    a(t)=\frac{\chi_0\Omega_{k,0}}{\Omega_{r,0}}+\frac{H_0^2\Omega_{r,0}}{4\chi_0}(t-t_k)^2,
\end{equation}
which clearly shows that the scale factor reached a minimum at $t_k$, meaning the universe had undergone a cosmic bounce. It is a peculiar feature of this model, however, that the bouncing solution is found in the case of negative spatial curvature. Bounce models commonly found in the literature generally require positive spatial curvature in order to avoid violations of the energy conditions for matter components~\cite{Peter:2001fy}. In other words, the negative energy contribution of positive spatial curvature (for which, we recall, $\Omega_k<0$) is what reverses the attractive nature of gravity. In our case, however, the non-minimal coupling drastically changes the role of spatial curvature: whenever the non-minimal coupling is relevant, it is negative spatial curvature that contributes repulsively to gravity, allowing a contracting universe to bounce. We note that a bouncing universe with positive spatial curvature could be found if the non-minimally coupled dark matter were a relativistic fluid. Nonetheless, we leave this case to Appendix~\ref{app|rel}, as it requires unnatural changes to the parameters of the theory.

While the natural emergence of an accelerated expansion phase and a cosmological bounce is appealing, these phenomena come with important caveats. In particular, the presence of derivatives of the energy density in the effective stress–energy tensor suggests that high-frequency perturbations could be amplified relative to low-frequency ones, thus distorting the nearly scale-invariant power spectrum of cosmological perturbations. A detailed analysis of perturbations is therefore necessary before this accelerated expansion phase can be regarded as a viable inflationary epoch, and such an analysis will be presented in a sequel to this work. Furthermore, the extreme energies at which the bounce occurs suggest that the fluid approximation may break down, requiring a more fundamental description of DM. Nonetheless, modified Friedmann equations can arise as effective descriptions of gravity even in purely quantum gravity regimes, leading to bouncing cosmologies~\cite{Sotiriou:2008ya,Barragan:2009sq}. Moreover, one can argue that a fluid description is appropriate when the mean free path is smaller than the characteristic length scale of the system, a condition that holds even more robustly in the high-density regime. Our results indicate that, in general, non-minimally coupled DM tends to interact with spatial curvature, potentially making it the dominant component in the very early Universe and leading to a cosmological bounce due to the gravitationally repulsive effect it can produce (since curvature is not constrained by energy conditions and can naturally take both positive and negative values). A bounce might then be a generic feature of non-minimally coupled models, and could provide an alternative to inflation as a mechanism for generating a nearly scale-invariant power spectrum of cosmological perturbations~\cite{Finelli:2001sr}.

\section{Conclusions}

In this paper, we described the impact of a non-minimal coupling between dark matter and gravity on the background cosmological evolution. The coupling was considered at the fluid level to remain agnostic about its fundamental origin or the nature of dark matter, and was constructed in a perturbative-inspired manner by coupling a pressureless stress-energy tensor to the Einstein curvature tensor. The resulting action requires the introduction of a new length scale, which was fixed by requiring that the interaction cross-section remains subdominant compared to Standard Model interactions. In a previous paper, we showed that this value of the length scale gives rise to self-gravitating dark matter configurations with masses compatible with those of supermassive black hole seeds.

The Friedmann and Raychaudhuri equations are modified by the non-minimal coupling, while the conservation equations remain unchanged. It is thus possible to infer the cosmic expansion directly from the evolution of the energy densities. We found that the non-minimal coupling drives a phase of accelerated expansion in the early universe, which, for the chosen value of the length scale, ends just before neutrino decoupling. We verified that this accelerated phase lasts long enough to solve the horizon problem, and could therefore serve as an inflationary epoch that does not require the introduction of additional fields.

Within this model the origin of the Universe depends strongly on spatial curvature. While positive or zero curvature always leads to an initial singularity, negative spatial curvature allows for the possibility of a cosmological bounce. The non-minimal coupling to spatial curvature makes it the dominant component in the very early Universe and causes it to reverse its usual effect: negative curvature reduces the expansion rate, while positive curvature increases it, opposite to what happens in standard cosmology. The scale factor at which the bounce occurs depends on the chosen length scale and the current value of spatial curvature. For a curvature value close to current observational bounds, the bounce take place precisely at the Planck epoch.

Since models of the early universe are now severely constrained by precision CMB measurements, it becomes necessary to predict the primordial power spectrum of linear perturbations. However, the power-law behavior of the background prevents the use of slow-roll approximations. We are therefore conducting a full analysis of cosmological perturbations, which will be presented in a sequel to this paper.

\acknowledgments

This work was partially funded from the projects: `Data Science methods for MultiMessenger Astrophysics \& Multi-Survey Cosmology' funded by the Italian Ministry of University and Research, Programmazione triennale 2021/2023 (DM n.2503 dd. 9 December 2019), Programma Congiunto Scuole; EU H2020-MSCA-ITN-2019 n. 860744 \textit{BiD4BESt: Big Data applications for black hole Evolution STudies}; Italian Research Center on High Performance Computing Big Data and Quantum Computing (ICSC), project funded by European Union - NextGenerationEU - and National Recovery and Resilience Plan (NRRP) - Mission 4 Component 2 within the activities of Spoke 3 (Astrophysics and Cosmos Observations);  European Union - NextGenerationEU under the PRIN MUR 2022 project n. 20224JR28W `Charting unexplored avenues in Dark Matter'; INAF GO-GTO Normal 2023 funding scheme with the project "Serendipitous H-ATLAS-fields Observations of Radio Extragalactic Sources (SHORES)"; INAF Large Grant 2022 project "MeerKAT and LOFAR Team up: a Unique Radio Window on Galaxy/AGN co-Evolution; INAF Large GO 2024 project "MeerKAT and Euclid Team up: Exploring the galaxy-halo connection at cosmic noon".

\appendix
\section{Cosmology with a non-minimally coupled relativistic fluid}\label{app|rel}

For the sake of completeness, we consider the case in which dark matter is modeled as a relativistic fluid rather than as dust. This approach accounts for the possibility that the non-minimal coupling becomes relevant at sufficiently early times, when dark matter particles have relativistic velocities.
The Friedmann equation~\eqref{eq|friedmannchi} in this case becomes
\begin{equation}\label{eq|friedrelat}
\left(\frac{H}{H_0}\right)^2 = \frac{\Omega_{b,0}(1+z)^3 + \Omega_{rel,0}(1+z)^4 + \Omega_{\Lambda,0} + \Omega_{k,0}(1+z)^2\,(1+\epsilon\bar{\chi}_0(1+z)^4)}{1+\epsilon\bar{\chi}_0(1+z)^4},    
\end{equation}
where $\bar{\chi}_0=\frac{16}{3}\pi G L^2\rho_{\rm DM}$, $\Omega_{rel,0}=\Omega_{r,0}+\Omega_{DM,0}$, and we leave the polarity parameter $\epsilon$ unspecified. Three major differences appear from equation~\eqref{eq|friedrelat}:
\begin{itemize}
    \item[-] to avoid a curvature singularity at a finite value of $z$, we must set $\epsilon=+1$;
    \item[-] the end of the NMC phase is delayed, with $\log(1+\bar{z}_{NMC})=\frac{3}{4}\log(1+z_{NMC})$;
    \item[-] curvature is non-minimally coupled as $\Omega_k(1+\bar{\chi})$ rather than $\Omega_k(1-\chi)$.
\end{itemize}
The first two points pose serious challenges to the model. The first suggests that the non-minimal coupling term in the action would need to change sign as the equation of state parameter transitions from $w=0$ to $w=1/3$, yet there is no natural way to implement such a change.  The second indicates that the end of the non-minimal coupling–driven phase occurs much later in cosmic history, potentially spoiling standard cosmological predictions and contradicting its own assumptions, since dark matter is expected to be non-relativistic at later times. The only possible workaround is to assume that the length scale $L$ decreases by several orders of magnitude as $w$ changes from $0$ to $1/3$, but again, no natural mechanism exists to justify this transition.

Nonetheless, if we assume that such mechanisms exist, the resulting evolution is quite intriguing. When the non-minimal coupling is significant, i.e. when $\bar{\chi}\gg1$, equation~\eqref{eq|friedrelat} becomes 
\begin{equation}\label{eq|friedrelatnmc}
\left(\frac{H}{H_0}\right)^2 = \frac{\Omega_{rel,0}}{\bar{\chi}_0}+\Omega_{k,0}(1+z)^2.    
\end{equation}
Whenever $\Omega_{k,0}(1+z)^2\ll1$, the universe undergoes a phase of exponential expansion, as predicted by standard inflationary models. Moreover, the cosmological bounce occurs with positive spatial curvature $k>0$ (and $\Omega_k<0$), like in conventional cosmological bounce scenarios~\cite{Peter:2001fy,Novello:2008ra}.

\bibliographystyle{unsrt}
\bibliography{biblio}

\begin{thebibliography}{10}

\bibitem{Bull:2015stt}
Philip Bull et~al.
\newblock {Beyond $\Lambda$CDM: Problems, solutions, and the road ahead}.
\newblock {\em Phys. Dark Univ.}, 12:56--99, 2016.

\bibitem{Nojiri:2017ncd}
S.~Nojiri, S.~D. Odintsov, and V.~K. Oikonomou.
\newblock {Modified Gravity Theories on a Nutshell: Inflation, Bounce and Late-time Evolution}.
\newblock {\em Phys. Rept.}, 692:1--104, 2017.

\bibitem{Carroll:2000fy}
Sean~M. Carroll.
\newblock {The Cosmological constant}.
\newblock {\em Living Rev. Rel.}, 4:1, 2001.

\bibitem{SupernovaCosmologyProject:1998vns}
S.~Perlmutter et~al.
\newblock {Measurements of $\Omega$ and $\Lambda$ from 42 High Redshift Supernovae}.
\newblock {\em Astrophys. J.}, 517:565--586, 1999.

\bibitem{Bertone:2004pz}
Gianfranco Bertone, Dan Hooper, and Joseph Silk.
\newblock {Particle dark matter: Evidence, candidates and constraints}.
\newblock {\em Phys. Rept.}, 405:279--390, 2005.

\bibitem{Lapi:2023plb}
Andrea Lapi, Lumen Boco, Marcos~M. Cueli, Balakrishna~S. Haridasu, Tommaso Ronconi, Carlo Baccigalupi, and Luigi Danese.
\newblock {Little Ado about Everything: \ensuremath{\eta}CDM, a Cosmological Model with Fluctuation-driven Acceleration at Late Times}.
\newblock {\em Astrophys. J.}, 959(2):83, 2023.

\bibitem{Famaey:2011kh}
Benoit Famaey and Stacy McGaugh.
\newblock {Modified Newtonian Dynamics (MOND): Observational Phenomenology and Relativistic Extensions}.
\newblock {\em Living Rev. Rel.}, 15:10, 2012.

\bibitem{Allen2011}
Steven~W. {Allen}, August~E. {Evrard}, and Adam~B. {Mantz}.
\newblock {Cosmological Parameters from Observations of Galaxy Clusters}.
\newblock {\em ARA\&A}, 49(1):409--470, September 2011.

\bibitem{Rubin1980}
V.~C. {Rubin}, Jr. {Ford}, W.~K., and N.~{Thonnard}.
\newblock {Rotational properties of 21 SC galaxies with a large range of luminosities and radii, from NGC 4605 (R=4kpc) to UGC 2885 (R=122kpc).}
\newblock {\em ApJ}, 238:471--487, June 1980.

\bibitem{Planck}
N.~Aghanim et~al.
\newblock {Planck 2018 results. VI. Cosmological parameters}.
\newblock {\em Astron. Astrophys.}, 641:A6, 2020.
\newblock [Erratum: Astron.Astrophys. 652, C4 (2021)].

\bibitem{Paraficz2016}
Paraficz~D. et~Al.
\newblock {The Bullet cluster at its best: weighing stars, gas, and dark matter}.
\newblock {\em A\&A}, 594:A121, October 2016.

\bibitem{DiValentino:2021izs}
Eleonora Di~Valentino, Olga Mena, Supriya Pan, Luca Visinelli, Weiqiang Yang, Alessandro Melchiorri, David~F. Mota, Adam~G. Riess, and Joseph Silk.
\newblock {In the realm of the Hubble tension\textemdash{}a review of solutions}.
\newblock {\em Class. Quant. Grav.}, 38(15):153001, 2021.

\bibitem{DESI:2024mwx}
A.~G. Adame et~al.
\newblock {DESI 2024 VI: cosmological constraints from the measurements of baryon acoustic oscillations}.
\newblock {\em JCAP}, 02:021, 2025.

\bibitem{Guth:1980zm}
Alan~H. Guth.
\newblock {The Inflationary Universe: A Possible Solution to the Horizon and Flatness Problems}.
\newblock {\em Phys. Rev. D}, 23:347--356, 1981.

\bibitem{Albrecht:1982wi}
Andreas Albrecht and Paul~J. Steinhardt.
\newblock {Cosmology for Grand Unified Theories with Radiatively Induced Symmetry Breaking}.
\newblock {\em Phys. Rev. Lett.}, 48:1220--1223, 1982.

\bibitem{Starobinsky:1980te}
Alexei~A. Starobinsky.
\newblock {A New Type of Isotropic Cosmological Models Without Singularity}.
\newblock {\em Phys. Lett. B}, 91:99--102, 1980.

\bibitem{Kofman:1994rk}
Lev Kofman, Andrei~D. Linde, and Alexei~A. Starobinsky.
\newblock {Reheating after inflation}.
\newblock {\em Phys. Rev. Lett.}, 73:3195--3198, 1994.

\bibitem{Peccei:1977hh}
R.~D. Peccei and Helen~R. Quinn.
\newblock {CP Conservation in the Presence of Instantons}.
\newblock {\em Phys. Rev. Lett.}, 38:1440--1443, 1977.

\bibitem{Duffy:2009ig}
Leanne~D. Duffy and Karl van Bibber.
\newblock {Axions as Dark Matter Particles}.
\newblock {\em New J. Phys.}, 11:105008, 2009.

\bibitem{Chadha-Day:2021szb}
Francesca Chadha-Day, John Ellis, and David J.~E. Marsh.
\newblock {Axion dark matter: What is it and why now?}
\newblock {\em Sci. Adv.}, 8(8):abj3618, 2022.

\bibitem{Carr:2016drx}
Bernard Carr, Florian Kuhnel, and Marit Sandstad.
\newblock {Primordial Black Holes as Dark Matter}.
\newblock {\em Phys. Rev. D}, 94(8):083504, 2016.

\bibitem{Green:2024bam}
Anne~M. Green.
\newblock {Primordial black holes as a dark matter candidate - a brief overview}.
\newblock {\em Nucl. Phys. B}, 1003:116494, 2024.

\bibitem{Jungman:1995df}
Gerard Jungman, Marc Kamionkowski, and Kim Griest.
\newblock {Supersymmetric dark matter}.
\newblock {\em Phys. Rept.}, 267:195--373, 1996.

\bibitem{XENON1T}
Aprile~E. et~Al.
\newblock Dark matter search results from a one ton-year exposure of xenon1t.
\newblock {\em Phys. Rev. Lett.}, 121:111302, Sep 2018.

\bibitem{PANDAX}
Meng~Yue et. Al.
\newblock Dark matter search results from the pandax-4t commissioning run.
\newblock {\em Phys. Rev. Lett.}, 127:261802, Dec 2021.

\bibitem{LZ}
Aalbers~J. et~Al.
\newblock First dark matter search results from the lux-zeplin (lz) experiment.
\newblock {\em Physical Review Letters}, 131(4), July 2023.

\bibitem{Birrell:1982ix}
N.~D. Birrell and P.~C.~W. Davies.
\newblock {\em {Quantum Fields in Curved Space}}.
\newblock Cambridge Monographs on Mathematical Physics. Cambridge University Press, Cambridge, UK, 1982.

\bibitem{Hertzberg:2010dc}
Mark~P. Hertzberg.
\newblock {On Inflation with Non-minimal Coupling}.
\newblock {\em JHEP}, 11:023, 2010.

\bibitem{Bettoni14}
Dario {Bettoni}, Mattia {Colombo}, and Stefano {Liberati}.
\newblock {Dark matter as a Bose-Einstein Condensate: the relativistic non-minimally coupled case}.
\newblock {\em JCAP}, 2014(2):004, February 2014.

\bibitem{Bettoni11}
Dario {Bettoni}, Stefano {Liberati}, and Lorenzo {Sindoni}.
\newblock {Extended {\ensuremath{\Lambda}}CDM: generalized non-minimal coupling for dark matter fluids}.
\newblock {\em JCAP}, 2011(11):007, November 2011.

\bibitem{Bettoni12}
Dario {Bettoni}, Valeria {Pettorino}, Stefano {Liberati}, and Carlo {Baccigalupi}.
\newblock {Non-minimally coupled dark matter: effective pressure and structure formation}.
\newblock {\em JCAP}, 2012(7):027, July 2012.

\bibitem{Bettoni15}
Dario {Bettoni} and Stefano {Liberati}.
\newblock {Dynamics of non-minimally coupled perfect fluids}.
\newblock {\em JCAP}, 2015(8):023--023, August 2015.

\bibitem{Gandolfi21}
Giovanni {Gandolfi}, Andrea {Lapi}, and Stefano {Liberati}.
\newblock {Self-gravitating Equilibria of Non-minimally Coupled Dark Matter Halos}.
\newblock {\em ApJ}, 910(1):76, March 2021.

\bibitem{Gandolfi22}
Giovanni {Gandolfi}, Andrea {Lapi}, and Stefano {Liberati}.
\newblock {Empirical Evidence of Nonminimally Coupled Dark Matter in the Dynamics of Local Spiral Galaxies?}
\newblock {\em ApJ}, 929(1):48, April 2022.

\bibitem{Gandolfi23}
Giovanni {Gandolfi}, Balakrishna~S. {Haridasu}, Stefano {Liberati}, and Andrea {Lapi}.
\newblock {Looking for Traces of Nonminimally Coupled Dark Matter in the X-COP Galaxy Clusters Sample}.
\newblock {\em ApJ}, 952(2):105, August 2023.

\bibitem{Benetti_2025}
Francesco Benetti, Andrea Lapi, Samuele Silveravalle, and Stefano Liberati.
\newblock Ultra-compact objects of non-minimally coupled dark matter.
\newblock {\em Journal of Cosmology and Astroparticle Physics}, 2025(03):029, mar 2025.

\bibitem{CMS:2018ccg}
Albert~M. Sirunyan et~al.
\newblock {Measurement of differential cross sections for Z boson pair production in association with jets at $\sqrt{s} =$ 8 and 13 TeV}.
\newblock {\em Phys. Lett. B}, 789:19--44, 2019.

\bibitem{CMS:2020fqz}
Albert~M Sirunyan et~al.
\newblock {Evidence for electroweak production of four charged leptons and two jets in proton-proton collisions at $\sqrt {s}$ = 13 TeV}.
\newblock {\em Phys. Lett. B}, 812:135992, 2021.

\bibitem{Johnson:2012cw}
Jarrett~L. Johnson, Daniel~J. Whalen, Hui Li, and Daniel~E. Holz.
\newblock {Supermassive Seeds for Supermassive Black Holes}.
\newblock {\em Astrophys. J.}, 771:116, 2013.

\bibitem{Bekenstein93}
Jacob~D. {Bekenstein}.
\newblock {Relation between physical and gravitational geometry}.
\newblock {\em PRD}, 48(8):3641--3647, October 1993.

\bibitem{Bettoni:2013diz}
Dario Bettoni and Stefano Liberati.
\newblock {Disformal invariance of second order scalar-tensor theories: Framing the Horndeski action}.
\newblock {\em Phys. Rev. D}, 88:084020, 2013.

\bibitem{haghani2023variation}
Zahra Haghani, Tiberiu Harko, and Shahab Shahidi.
\newblock {The first variation of the matter energy{\textendash}momentum tensor with respect to the metric, and its implications on modified gravity theories}.
\newblock {\em Phys. Dark Univ.}, 44:101448, 2024.

\bibitem{Dubovsky:2011sj}
Sergei Dubovsky, Lam Hui, Alberto Nicolis, and Dam~Thanh Son.
\newblock {Effective field theory for hydrodynamics: thermodynamics, and the derivative expansion}.
\newblock {\em Phys. Rev. D}, 85:085029, 2012.

\bibitem{Ballesteros:2012kv}
Guillermo Ballesteros and Brando Bellazzini.
\newblock {Effective perfect fluids in cosmology}.
\newblock {\em JCAP}, 04:001, 2013.

\bibitem{Ballesteros:2014sxa}
Guillermo Ballesteros.
\newblock {The effective theory of fluids at NLO and implications for dark energy}.
\newblock {\em JCAP}, 03:001, 2015.

\bibitem{Ballesteros:2016gwc}
Guillermo Ballesteros, Denis Comelli, and Luigi Pilo.
\newblock {Massive and modified gravity as self-gravitating media}.
\newblock {\em Phys. Rev. D}, 94(12):124023, 2016.

\bibitem{Berezhiani:2015bqa}
Lasha Berezhiani and Justin Khoury.
\newblock {Theory of dark matter superfluidity}.
\newblock {\em Phys. Rev. D}, 92:103510, 2015.

\bibitem{Peter:2001fy}
Patrick Peter and Nelson Pinto-Neto.
\newblock {Has the universe always expanded?}
\newblock {\em Phys. Rev. D}, 65:023513, 2002.

\bibitem{Sotiriou:2008ya}
Thomas~P. Sotiriou.
\newblock {Covariant Effective Action for Loop Quantum Cosmology from Order Reduction}.
\newblock {\em Phys. Rev. D}, 79:044035, 2009.

\bibitem{Barragan:2009sq}
Carlos Barragan, Gonzalo~J. Olmo, and Helios Sanchis-Alepuz.
\newblock {Bouncing Cosmologies in Palatini f(R) Gravity}.
\newblock {\em Phys. Rev. D}, 80:024016, 2009.

\bibitem{Finelli:2001sr}
Fabio Finelli and Robert Brandenberger.
\newblock {On the generation of a scale invariant spectrum of adiabatic fluctuations in cosmological models with a contracting phase}.
\newblock {\em Phys. Rev. D}, 65:103522, 2002.

\bibitem{Novello:2008ra}
M.~Novello and S.~E.~Perez Bergliaffa.
\newblock {Bouncing Cosmologies}.
\newblock {\em Phys. Rept.}, 463:127--213, 2008.

\end{thebibliography}
\end{document}